\tolerance   10000
\magnification = 1150
\baselineskip=\normalbaselineskip
\advance\hoffset by  .4in
\advance\hsize   by -.97in
\advance\vsize   by .22in
\font\timesbold=cmb10 scaled  1500

\font\timesboldeleven=cmb10 scaled 1000

\font\figures=cmr10 scaled 800
\font\caption=cmb10 scaled 800
\let\origeqno=\eqno
\def\eqno(#1){\origeqno (\rm #1)}
\input epsf

\topskip 6.2pc
\vbox{
\noindent
In memory of J. C. Huang, late professor of Physics
at the University of Missouri-Columbia.
\vskip 2in
\noindent
{\timesbold {On the Asymptotic Character of Electromagnetic Waves
in a Friedmann-Robertson-Walker Universe}}}
\vskip 30pt
\noindent
{\timesboldeleven {Nader Haghighipour$^{1}$}}
\vskip  29pt
\hbox{
\hskip 0.65in
\advance\hsize   by -.72in
\vbox{
\baselineskip=0.8\normalbaselineskip
\noindent
\line{\hrulefill}
\figures Asymptotic properties of electromagnetic waves
are studied within the context of
Friedmann-Robertson-Walker (FRW) cosmology.
Electromagnetic fields are considered as small perturbations on the
background spacetime and Maxwell's equations
are solved for all three cases of flat, closed and open FRW universes.
The asymptotic character of these solutions
are investigated and their relevance 
to the problem of cosmological tails of electromagnetic
waves is discussed.
\line{\hrulefill}
}}
\vskip 40pt
\noindent
{\timesboldeleven {1. INTRODUCTION}}
\vskip 10pt

It is well understood that while propagating,
similar to any traveling wave,
electromagnetic radiation interacts 
with the curvature of the spacetime [1]. Such 
interactions affect the propagation of these waves
and cause them to scatter. The scattered waves, or tails,
manifest themselves as partial backscattering when the source
of the curvature is a localized object. However,
when propagating throughout the universe, where
the curvature exists everywhere, the scattering of 
electromagnetic waves due to the interaction with the background 
curvature occurs throughout the space and at all times. 
The purpose of this paper is to understand how the background
curvature of spacetime affects the propagation and properties of 
electromagnetic waves, particularly at large distances.

\line{\hrulefill \hskip 4.16in}
\vskip -5pt
\noindent
$^{\figures {1}}${\figures Department of Terrestrial Magnetism,
Carnegie Institution of Washington, 5241 Broad Branch Road, NW,
Washington, DC 20015; e-mail: nader@dtm.ciw.edu}
\vskip 60pt

Studies of this nature have been done by Faraoni and Sonego[2,3]
on the analysis of the propagation of scalar fields in
FRW spacetime. In those papers, the authors have presented
analytical solutions to the Klein-Gordon equation with non-minimal
coupling and by studying the asymptotic character of these
solutions, they have shown that standard calculations of the 
reflection and transmission coefficients are not applicable to 
this case. Scialom and Philippe [4] have also studied the asymptotic
behavior of scalar fields near the singular points of
Einstein-Klein-Gordon equation in a flat FRW universe.
They determined the
singular points of Einstein-Klein-Gordon equation
and by analytically analyzing the asymptotic behavior 
of complex scalar fields, showed that these fields,
similar to real scalar fields, support the existence
of inflationary stage in a flat FRW universe.
For vector fields,
Noonan [5,6,7] studied the propagation of electromagnetic
fields in a curved spacetime and showed that,
while propagating in a conformally flat universe,
components of the Faraday tensor stay tail-free.
In this paper, propagation of electromagnetic waves
in the FRW spacetime is studied and the explicit 
asymptotic character of these field in all three
cases of flat, closed, and open universes will be 
discussed.

To investigate the
asymptotic character of electromagnetic fields
one has to calculate the electric and magnetic
components of these fields as measured by a standard observer, and
study their properties at large distances. Here
electromagnetic fields are considered
as small perturbations on the background curvature, and
solutions of Maxwell's equations are presented 
for all three cases of flat, closed, and open FRW universes.
Asymptotic properties of electromagnetic fields are then
investigated by studying these solutions at large distances.

Solutions to Maxwell's equations in a FRW universe
have been presented by different authors [8-11]. 
Deng and Mannheim [12]
have presented solutions to Maxwell's equations 
by solving these equations for spherical components
of the electric and magnetic fields, separately. 
More recently, Mankin et al [13,14,15] have 
presented exact and approximate solutions to 
electromagnetic wave equations in a curved spacetime,
based on the method proposed by Hadamard [16] and by
using a higher-order Green's function for 
the wave equation.
I adopt a very helpful method due to Skrotskii [17]
who realized that electromagnetic field equations 
in a curved spacetime can be written
in a non-covariant form formally equivalent to
Maxwell's equations in a macroscopic medium in flat spacetime.
The electric and magnetic properties of this medium are
tied to the background curvature of spacetime [17,18].
This method has been used by Mashhoon to solve
Maxwell's equations in a closed expanding FRW universe [19].

The plan of this paper is as follows. Maxwell's equations are 
discussed in section 2. In section 3, solutions to
Maxwell's equations are presented for all three cases of flat,
closed and open FRW universes. Section 4 has to do with the
study of the asymptotic character of electromagnetic fields,
and section 5 concludes 
this study by reviewing the results and discussing their applications.

The metric convention $g_{\alpha\beta}=(+,-,-,-)$ 
with $d{s^2}={g_{\alpha\beta}}d{x^\alpha}d{x^\beta}$ 
is used throughout this paper. The Greek indices will
indicate sums over 0,1,2 and 3 while the Latin indices will sum over
1,2 and 3. The units used in this paper have been chosen such that
$\hbar=c=1$ where $c$ is the speed of light.
\vskip  25pt
\noindent
{\timesboldeleven {2. MAXWELL'S EQUATIONS IN FRW SPACETIME}}
\vskip  10pt

In a curved spacetime with metric $g_{\alpha\beta}$, 
the source-free Maxwell's equations are given by
$$
{\bigl[(-g)^{1/2}F^{\alpha\beta}\bigl]_,}_{\,\beta} = 0\>,
\eqno (1)
$$
and
$$
F_{\alpha\beta,\gamma}\,+\,F_{\gamma\alpha,\beta}\,+\,
F_{\beta\gamma,\alpha}\,=\,0\>,
\eqno  (2)
$$
where $F_{\alpha\beta}$ is the electromagnetic field tensor,
$g={\rm det}(g_{\alpha\beta})$, and (,) denotes an ordinary
differentiation. In accordance with Skrotskii's
formalism [17], a background inertial frame is introduced
with Cartesian coordinates in which the electric and
magnetic fields are defined using the decompositions
${F_{\alpha\beta}} \to ({\vec E},{\vec B})$,  and
${\sqrt{-g}}{F^{\alpha\beta}} \to (-{\vec D},{\vec H})$.
That is,
$$
E_a=F_{a0}\qquad,\qquad D_a=(-g)^{1/2} F^{0a}\>,
\eqno  (3)
$$
and
$$
B_a={1\over 2}\, \epsilon_{abc} F_{bc}\qquad,\qquad
H_a={1\over 2}\, \epsilon_{abc} (-g)^{1\over 2} F^{bc}\>,
\eqno  (4)
$$
\noindent
where $\epsilon_{abc}$ is the three-dimensional Levi-Civita
symbol. Using these quantities, 
equations (1) and (2) can be written as [18]
$$
D_{a,a} = B_{a,a}= 0\>,
\eqno  (5)
$$
and
$$
-D_{a,0}+\epsilon_{abc} H_{c,b}=B_{a,0}+\epsilon_{abc} E_{c,b}=0\>.
\eqno  (6)
$$

Equations (5) and (6) are formally equivalent to
the electromagnetic field equations
in a medium in flat spacetime with a dielectric constant 
$\varepsilon_{ab}$ and a permutivity $\mu_{ab}$ given by
$$
\varepsilon_{ab}=\mu_{ab}=-(-g)^{1/2}\>{g^{ab}\over{g_{_{00}}}}\>.
\eqno(7)
$$
\noindent
In this auxiliary medium and with the assumed Cartesian coordinate system, 
$({\vec E},{\vec D})$ represent the electric
fields and $({\vec B},{\vec H})$ are their corresponding magnetic fields.
These vector fields are related via constitutive relations [19]
$$
D_a = \varepsilon_{ab} E_b - (\vec G\times \vec H)_a \>,
\eqno(8)
$$
$$
B_a = \mu_{ab} H_b + (\vec G \times\vec E)_a \>,
\eqno(9)
$$
\noindent
where $G_a=-{g_{{_0}{_a}}/{g_{_{00}}}}$.

The above-mentioned formalism is applicable to any curved spacetime. 
In this study the curved spacetime of interest is 
FRW with a metric given by
\vskip 2pt
$$
{(ds)}^2 ={(dt)}^2 -{{{\cal S}^2}(t)}{R_0^2}\,\biggl\{{(d{\cal R})^2 \over 
{(1-k{\cal R}^2)}}+{\cal R}^2(d\theta)^2
+{\cal R}^2\sin^2\theta(d\phi)^2 \biggl\}\>.
\eqno(10)
$$
\vskip 3pt
\noindent
In equation (10), $R_0$ is the radius of the model universe
at some epoch $t_0$. The expansion parameter of this model
universe is given by ${\cal S}(t)$ such that ${\cal S}({t_0})=1$. 
The product ${\cal S}(t){R_0}$ in equation (10)
is the cosmic scale factor, and
$k=-1,0,+1$ corresponding to open, flat, and closed 
universes, respectively. For $k=+1$, ${\cal R}$ ranges from 0 to 1.

Introducing ${x_1}\,,{x_2}$ and ${x_3}$ given by 
$r\sin \theta \cos \phi,\, r \sin \theta \sin \phi$
and $ r \cos \theta$, respectively,
the line element (10) can be written as
$$
(ds)^2 = {{\cal C}^2}(\eta) \biggl\{ (d\eta)^2 - \big[{f_k}(r)\big]^2 
(\delta_{ww'} dx^w dx^{w'}) \biggl\}\>.
\eqno (11)
$$
\noindent
In this equation, $\eta$, the {\it conformal time},
is given by $dt={\cal S}(t)d\eta$, and 
${\cal C}(\eta)={\cal S}(t)$. The function ${f_k}(r)$
in equation (11) is equal to
\vskip 5pt
$$
{f_k}(r) = \biggl[1+k{\Bigl({r\over{2{R_0}}}\Bigr)^2}
\biggr]^{-1}\>,
\eqno  (12)
$$
\noindent
where $r$ is defined via 
$$
{\cal R}= {r\over {R_0}} \biggl[1+k
{\Bigl({r\over{2{R_0}}}\Bigr)^2}\biggr]^{-1}\>.
\eqno  (13)
$$

In terms of the new coordinates $(\eta,{x_1},{x_2},{x_3})$,  
the dielectric constant $\varepsilon_{ab}$ and 
permutivity $\mu_{ab}$ are given by
$$
\varepsilon_{ab} = \mu_{ab} = {f_k}(r)\delta_{ab}\>.
\eqno (14)
$$
Also in this new coordinate system, ${G_a}=0$. 
Substituting these quantities in equations (5) and (6), 
one can show that Maxwell's equations can be written as [12]
$$
-i\vec \nabla \times \vec {\cal I} = {f_k}(r)
{\partial {\vec {\cal I}} \over {\partial \eta}}\>,
\eqno (15)
$$
\noindent
and
$$
\vec \nabla \cdot \Bigl[{f_k}(r) \vec {\cal I}\, \Bigl] = 0\>,
\eqno (16)
$$
\noindent
where ${\vec {\cal I}}={\vec E}+i{\vec H}$. 
Equations (15) and (16) represent the electromagnetic field 
equations for the line element (11).
It is evident from these equations that for the metric (11),
electromagnetic fields are independent of ${\cal S}(t)$. 
This is due to the explicit conformal invariance of the
formalism above, and implies that
in a universe with a background curvature given by
equation (11), the
physical parameters of electromagnetic fields vary adiabatically with the
changing {\it radius of the universe}.
Among the solutions of equations (15) and (16), 
only those that represent $\vec E$ and $\vec H$ 
at a conformal time $\eta$, corresponding to $t={\tilde \tau}$, are 
solutions to the Maxwell's 
equations in a FRW universe at time ${\tilde \tau}$. 
\vskip  20pt
\noindent
{\timesboldeleven {3. SOLUTIONS OF MAXWELL'S EQUATIONS}}
\vskip  10pt

The isotropy and homogeneity of FRW spacetime implies that it is
possible to find solutions to equations (15) and (16) for a
definite angular momentum $J$, and its component along the
$x_3$-axis, $M$. Denoting such solutions by
$\vec {\cal I}_{\!J\!M\!{\sigma_k}}({\vec r}\,,\,\eta)$,
one can write
$$
\vec {\cal I}_{\!J\!M\!{\sigma_k}}({\vec r}\,,\,\eta) =
\sum_{\lambda = e,m,0} {\cal I}_{\!J\!{\sigma_k}}^\lambda(r)\> 
\vec Y_{\!\!J\!M}^\lambda (\theta,\phi)\>\>{e^{-i{\sigma_k} \eta}}\>,
\eqno (17)
$$
\vskip  3pt
\noindent
where $\vec {\cal I}_{\!J\!M\!{\sigma_k}}({\vec r}\,,\,\eta)$ 
have been expanded in terms of the vector spherical harmonics
$\vec Y_{\!\!J\!M}^\lambda (\theta,\phi)$ [19,20].
In equation (17), $\sigma_k$ is a positive constant
and $\lambda=e,m,0$ representing different types of  
vector spherical harmonics [20].
For a definite parity and angular momentum,
the electric and magnetic components of
$\vec {\cal I}_{\!J\!M\!{\sigma_k}}({\vec r}\,,\,\eta)$
can be written as [21]
\vskip 10pt
$$\!\!\!\!\!\!\!\!\!\!\!\!\!\!\!\!\!\!\!\!\!\!\!\!\!\!\!\!\!\!\!\!
\!\!\!\!\!\!\!\!\!\!\!\!\!\!\!\!\!\!\!\!\!\!\!\!\!\!\!\!\!\!\!\!
\!\!\!\!\!\!\!
{\vec E}_{\!\!J\!M\!{\sigma_k}}^m (\vec r\,)=
{\vec H}_{\!\!J\!M\!{\sigma_k}}^e (\vec r\,)=
{\cal I}_{\!J\!{\sigma_k}}^m (r) {\vec Y}_{\!\!J\!M}^m (\theta,\phi)\>,
\eqno (18)
$$
\vskip 5pt
\noindent
and
\vskip 2pt
$$
{\vec E}_{\!\!J\!M\!{\sigma_k}}^e (\vec r\,)=
-{\vec H}_{\!\!J\!M\!{\sigma_k}}^m (\vec r\,)=
i\biggl[{\cal I}_{\!J\!{\sigma_k}}^e (r){\vec Y}_{\!\!J\!M}^e (\theta,\phi) + 
{\cal I}_{\!J\!{\sigma_k}}^0 (r){\vec Y}_{\!\!J\!M}^0 (\theta,\phi )\biggl]\>,
\eqno  (19)
$$
\vskip 7pt
\noindent
where from equations (15) and (16), 
${{\cal I}_{\!J\!{\sigma_k}}^\lambda}(r)$ are given by
$$\!\!\!\!\!\!\!\!\!\!\!\!\!\!\!\!\!\!\!\!\!\!\!\!\!\!\!\!\!\!\!\!
\!\!\!\!\!\!\!\!\!\!\!\!\!\!\!\!\!\!\!\!\!\!\!\!\!\!\!\!\!\!\!\!
\!\!\!\!\!\!\!\!\!\!
{\cal I}_{\!J\!{\sigma_k}}^e (r) = - {1\over {{\sigma_k} {f_k}(r)}}
\biggl({d\over {dr}}+
{1\over r}\biggr)\, {\cal I}_{\!J\!{\sigma_k}}^m (r)\>,
\eqno  (20)
$$
\vskip 1pt
$$\!\!\!\!\!\!\!\!\!\!\!\!\!\!\!\!\!\!\!\!\!\!\!\!\!\!\!\!\!\!\!\!
\!\!\!\!\!\!\!\!\!\!\!\!\!\!\!\!\!\!\!\!\!\!\!\!\!\!\!\!
{\cal I}_{\!J\!{\sigma_k}}^0 (r) = -{1 \over {{\sigma_k} r {f_k}(r)}}\,
\Bigl[J(J+1)\Bigl]^{1/2}\, {\cal I}_{\!J\!{\sigma_k}}^m (r)\>,
\eqno  (21)
$$
\vskip 1pt
$$\eqalign{
\Biggl[\biggl({d^2 \over {d{r^2}}}\,+\,{2\over r}\,{d\over {dr}}\biggl)\,
-\,{1\over {{f_k}(r)}}\,&{{d{{f_k}(r)}}\over {dr}}\, 
\biggl({1\over r}\,+\,{d\over {dr}}\biggl)\cr 
&-\,{J(J+1) \over {r^2}}\,+\,{\sigma_k^2}
{f_k}^2(r)\Biggr]\,{\cal I}_{\!J\!{\sigma_k}}^m (r)= 0 \>.\cr}
\eqno  (22)
$$
\vskip 2pt
\noindent
The most general solution to equations (15) and (16) is a 
superposition of the fields (18) and (19). That is,
\vskip 2pt
$$\!\!\!\!\!\!\!\!\!\!\!\!\!\!\!\!
{\vec E} ({\vec r},\eta) = \sum_{J,M,{\sigma_k}} \sum_{\lambda' = e,m}
\biggl[{a_{\!\!J\!M\!{\sigma_k}}^{\lambda'}} (\eta) 
{\vec E}_{\!\!J\!M\!{\sigma_k}}^{\lambda'} (\vec r) +
{a_{\!\!J\!M\!{\sigma_k}}^{\lambda' \ast}} (\eta)
{\vec E}_{\!\!J\!M\!{\sigma_k}}^{\lambda' \ast} (\vec r)\biggr]\>,
\eqno  (23)
$$
\noindent
and
\vskip 3pt
$$\!\!\!\!\!\!\!\!\!\!\!\!\!\!\!\!
{\vec H} (\vec r,\eta) = \sum_{J,M,{\sigma_k}} \sum_{\lambda' = e,m}
\biggl[a_{\!\!J\!M\!{\sigma_k}}^{\lambda'} (\eta) 
{\vec H}_{\!\!J\!M\!{\sigma_k}}^{\lambda'} (\vec r)+
a_{\!\!J\!M\!{\sigma_k}}^{\lambda' \ast} (\eta)
{\vec H}_{\!\!J\!M\!{\sigma_k}}^{\lambda' \ast} (\vec r)\biggr]\>,
\eqno  (24)
$$
\noindent
where $\ast$ indicates complex conjugation and 
from equation (15), $a_{\!\!J\!M\!\sigma}^{\lambda'} (\eta)$ satisfies
$$
{d\over {d\eta}} {a_{\!\!J\!M\!{\sigma_k}}^{\lambda'}} (\eta) = 
-i{\sigma_k} {a_{\!\!J\!M\!{\sigma_k}}^{\lambda'}} (\eta)\>.
\eqno  (25)
$$

In the following, 
solutions of the differential equations 
(20) to (22) are presented for 
all three cases of flat, closed and open FRW universes.
\vskip 3in
\noindent
{\timesboldeleven {3-a.$\>$ Flat Universe}}
\vskip 5pt

In a flat universe where $k=0$, equation (22) is written as
$$
\biggl[ {{d^2}\over {d{r^2}}}\,+\,{2\over r}\,
\Bigl({d\over {dr}}\Bigr)\,-\,
{{J(J+1)}\over {r^2}}\,+\,{\sigma_0^2}\biggr]\,
{{\cal I}_{_{\!J\!{\sigma_0}}}^m}(r)\,=\,0\>.
\eqno  (26)
$$
\noindent
The general solution of equation (26) can be written
in terms of Hankel functions $h^{(1,2)}$. That is,
$$
{{\cal I}_{_{\!J\!{\sigma_0}}}^m}(r)\,=\,A\,{h_J^{(1)}}(r{\sigma_0})\,+\,
B\,{h_J^{(2)}}(r{\sigma_0})\,,
\eqno (27)
$$
\noindent
where $A$ and $B$ are constant quantities. In equation (27),
$$
{h_J^{(1)}}(r{\sigma_0})={j_J}(r{\sigma_0})+i{n_J}(r{\sigma_0}),
\eqno (28)
$$
\noindent
and
$$
{h_J^{(2)}}(r{\sigma_0})={j_J}(r{\sigma_0})-i{n_J}(r{\sigma_0}),
\eqno (29)
$$
\noindent
where ${j_J}$ and $n_J$ 
are spherical Bessel and Neumann functions, respectively, and
$J=0,1,2,3,...$ . 
Because ${n_J}(r{\sigma_0})$ is singular at $r=0$, 
solutions of equation (26) are proportional only to
spherical Bessel functions. Using
$$
{\int_0^\infty}{j_\ell}(r')\,{j_{\ell'}}(r')\,dr'\,=\,
{\pi\over {2(2\ell+1)}}\,{\delta_{\ell\ell'}}\,,
\eqno (30)
$$
\noindent
these solutions can be written as
\vskip 2pt
$$
{{\cal I}_{_{\!J\!{\sigma_0}}}^m}(r)\,=\,\biggl[{{2{\sigma_0}(2J+1)}\over 
\pi}\biggr]^{1/2}\,{j_J}(r{\sigma_0})\>,
\eqno (31)
$$
where, ${{\cal I}_{_{\!J\!{\sigma_0}}}^m}(r{\sigma_0})$ 
has been normalized such that
$$
{\int_0^\infty}{{\cal I}_{_{\!J\!{\sigma_0}}}^m}(r{\sigma_0})
{{\cal I}_{_{\!J'\!{\sigma_0}}}^m}(r{\sigma_0})\,dr\,=\,
{\delta_{JJ'}}.
\eqno (32)
$$
\vskip 1pt
\centerline{
\epsfbox{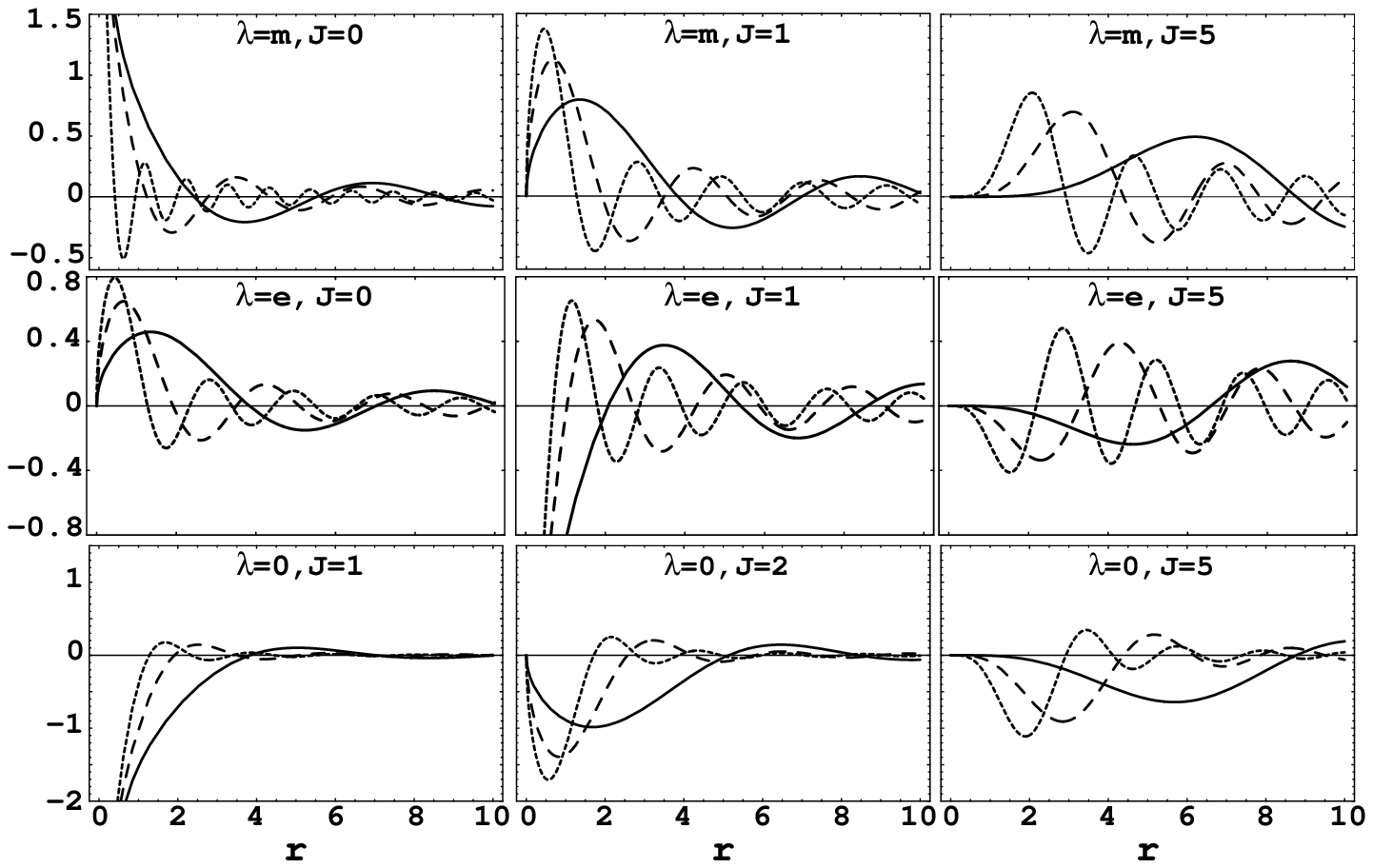}}
\vskip -1in
\vbox{
\baselineskip=0.9\normalbaselineskip
\noindent
\figures {\caption Figure 1.}$\>\>$
From top to bottom, 
graphs of ${{\cal I}_{_{\!J\!{\sigma_0}}}^\lambda}(r)$ for
$\lambda=m,e,0$, and 
for different values of $J$ and $\sigma_0$.
The solid line corresponds to ${\sigma_0}=1$, the dashed
line corresponds to ${\sigma_0}=2$ and the dotted line
corresponds to ${\sigma_0}=3$. Because 
${{\cal I}_{_{\!J\!{\sigma_0}}}^0}(r)=0$ for $J=0$,
the graphs of this function have been plotted for 
$J\geq 1$. It is also necessary to mention that the graphs of
$(\lambda=m , J=0)\,,(\lambda=e , J=1)$ and $(\lambda=0 , J=1)$
meet their corresponding vertical axes at values equal to
5, -8 and -9, respectively.
Note the different scales on the vertical axes.}
\vskip 30pt
\noindent
Substituting for  ${{\cal I}_{_{\!J\!{\sigma_0}}}^m}(r)$ 
from equation (31) in equations (20) and (21), one can show that 
$$
{{\cal I}_{_{\!J\!{\sigma_0}}}^e}(r)\,=
\,{\biggl[{{2{\sigma_0}}\over {\pi(2J+1)}}\biggr]^{1/2}}\,
\biggl[J\,{j_{_{J+1}}}(r{\sigma_0})-  
(J+1)\,{j_{_{J-1}}}(r{\sigma_0})\biggr],
\eqno  (33)
$$
\noindent
and
\vskip 2pt
$$\!\!\!\!\!\!\!\!\!\!\!\!\!\!\!\!\!\!\!\!\!\!\!\!\!\!\!\!\!\!\!\!
{{\cal I}_{_{\!J\!{\sigma_0}}}^0}(r)\,=
\,-\,{1\over r}\biggl[{{2J(J+1)(2J+1)}\over
{{\sigma_0}\pi}}\biggr]^{1/2}\,{j_{_J}}(r{\sigma_0})\>.
\eqno (34)
$$
\vskip 5pt
\noindent
Equations (31), (33) and (34) represent the $r$-dependence of 
the electric and magnetic fields given by equations (18) and (19).
Figure 1 shows these functions for different values of
$J$ and $\sigma_0$. From equation (34) one can see that
\vskip  40pt
{\centerline {
\epsfbox{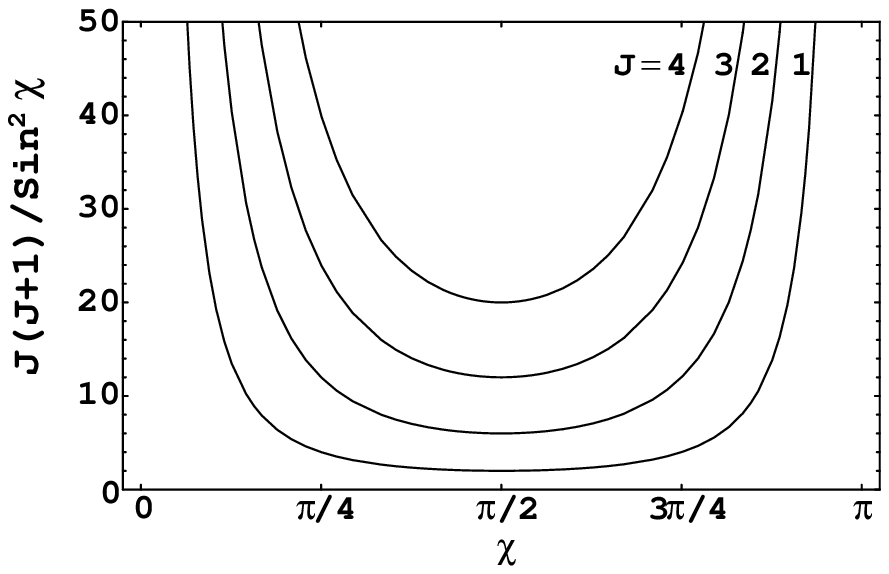}}}
\vbox{
\baselineskip=\normalbaselineskip
\vskip 45pt
\noindent
\figures {\caption Figure 2.}$\>\>$
Graph of $J(J+1)/{\sin^2}\chi$ against $\chi$ 
for $J=1,2,3,4$.}
\vskip 20pt
\noindent
${{\cal I}_{_{\!J\!{\sigma_0}}}^0}(r)$ vanishes when $J=0$. 
That means,
${\vec E}_{0\!M\!{\sigma_0}}^e (\vec r\,)$
and ${\vec H}_{0\!M\!{\sigma_0}}^m (\vec r\,)$ will be
pure $e$-type for all values of $\sigma_0$.
One can also see from  figure 1 that for any  
state of $\lambda$, when $J$ is constant, 
as expected, increasing $\sigma_0$ results in increasing
the frequency of the solutions of equation (26).
This figure also shows that among different states of
$\lambda,\,{{\cal I}_{_{\!J\!{\sigma_0}}}^0}(r)$ approaches
zero more rapidly implying that 
${{\cal I}_{_{\!J\!{\sigma_0}}}^m}(r)$ and
${{\cal I}_{_{\!J\!{\sigma_0}}}^e}(r)$  
will have dominating effects on the behavior of $\vec E$ and $\vec H$
fields at large distances.
\vskip 20pt
\noindent
{\timesboldeleven {3-b. Closed Universe}}
\vskip 5pt

In a closed FRW universe, $k=1$, and therefore 
${f_k}(r)={f_1}(r)=\qquad1/\{1+{[r/(2{R_0})]^2\}}$. Equation (22), 
in this case, is given by
$$
{{d^2}\over {d{\chi^2}}}{\Psi_{_{\!J\!{\omega_1}}}^m}(\chi)\,+\,
\biggl[{\omega_1^2}\,-\,
{{J(J+1)}\over{{\sin^2}\chi}}\biggr]\,
{\Psi_{_{\!J\!{\omega_1}}}^m}(\chi)\,=\,0\>,
\eqno  (35)
$$
\noindent
where ${\omega_1}={R_0}{\sigma_1}$,
$\chi=2{\tan^{-1}}[r/(2{R_0})]$, and
${\Psi_{_{\!J\!{\omega_1}}}^m}(\chi)=
{R_0}\,r\,{{\cal I}_{_{\!J\!{\sigma_1}}}^m}(r)$. 
In terms of the variable $\chi$, equation (35) resembles a 
Schr\"odinger equation with a {\it potential  function} given by 
$J(J+1)/{\sin^2}\chi$, and eigenvalues equal to $\omega_1^2$. 
Figure 2 shows the graph of this potential function for $J=1,2,3,4$.
It is evident from this figure that for all values of $J$, the
solutions of equation (35) must be finite 
for all values of $\chi \in [0,\pi]$. Using this
boundary condition,
the solution of equation(35) can be written as
\vskip 10pt
$$
{\Psi_{_{\!J\!{\omega_1}}}^m}(\chi)=
{2^{^{J+1}}}\,{{\omega_1} J!}\>
\Biggl[{({\omega_1}-J-1)!\over{({\omega_1} + J)!}}\Biggl]^{1/2}\>
(\sin \chi)^{J+1}\> C_{_{\!{\omega_1}\!-\!J\!-\!1}}^{^{J+1}} (\cos \chi)\>,
\eqno  (36)
$$
\noindent
where $J=1,2,3,...$ and ${\omega_1}=2,3,4,...$ 
representing frequencies  of different modes of
electromagnetic fields for the line element (11) [8,9,19]. 
In equation (36),
${C_\Upsilon^\upsilon} (\Theta)$ are Gegenbauer 
polynomials [22]
which are orthogonal over the interval $(-1,1)$ with the weight function
$W (\Theta) = (1-\Theta^2)^{\upsilon-{1/2}} \,\> ,\> \upsilon>-{1/2}$.
Orthogonality of ${C_{_{\!{\omega_1}\!-\!J\!-\!1}}^{^{J+1}}} (\cos \chi)$ 
requires ${\Psi_{_{\!J\!{\omega_1}}}^m}(\chi)$  to be normalized such that
$$
{\int_0^\pi} {\Psi_{_{\!J\!{\omega_1}}}^m}(\chi) 
{\Psi_{_{\!J\!{\omega'_1}}}^m}(\chi) d\chi = 
2\pi{\omega_1} \delta_{{\omega_1} {\omega'_1}}\>.
\eqno  (37)
$$
\vskip  3pt
\noindent
Using
$$
{d\over {d\Theta}}{C_\Upsilon^\upsilon}(\Theta)\,=\,
2\upsilon {C_{\Upsilon-1}^{\upsilon+1}}(\Theta)\,,
\eqno (38)
$$
\noindent
and the recurrence relation
$$
2\upsilon (1-{\Theta^2})\,{C_{\Upsilon-1}^{\upsilon+1}}(\Theta)\,=\,
(\Upsilon+2\upsilon -1){C_{\Upsilon-1}^\upsilon}(\Theta)\,-\,
\Upsilon \Theta {C_\Upsilon^\upsilon}(\Theta)\,,
\eqno (39)
$$
\noindent
and also from the definition of 
${\Psi_{_{\!J\!{\omega_1}}}^m}(\chi)$, and equations (20) and (21),
the $r$-dependence of the electric and magnetic fields
in a closed FRW universe are given by
$$\!
{{\cal I}_{_{\!J\!{\omega_1}}}^m}(\chi)\,=\,
{2^J}\,{{{\omega_1} J!}\over {R_0^2}}\>
\Biggl[{({\omega_1}-J-1)!\over{({\omega_1} + J)!}}\Biggl]^{1/2}\>
(1+\cos \chi)\,\,{\sin ^J}\chi\,\, 
C_{_{\!{\omega_1}\!-\!J\!-\!1}}^{^{J+1}} (\cos \chi)\>,
\eqno  (40)
$$
$$\!
\eqalign{
{{\cal I}_{_{\!J\!{\omega_1}}}^e}(\chi)=-{2^{J}}
{{J!}\over {{R_0^2}}}
&{\Biggl[{{({\omega_1}-J-1)!}\over {({\omega_1}+J)!}}\Biggr]^{1/2}}\>
(1+\cos \chi)\,\,{\sin ^J}\chi\cr
&\biggl[{\omega_1}\cos \chi\>{C_{_{{\omega_1}-J-1}}^{^{J+1}}}(\cos \chi)+
({\omega_1}+J)\,{C_{_{{\omega_1}-J-2}}^{^{J+1}}}(\cos \chi)\biggr]\,,\cr}
\eqno (41)
$$
$$\!
\eqalign{
{{\cal I}_{_{\!J\!{\omega_1}}}^0}(\chi)=-{2^{J}}
{{J!}\over {{R_0^2}}}
&{\Biggl[{{J(J+1)({\omega_1}-J-1)!}\over 
{({\omega_1}+J)!}}\Biggr]^{1/2}}\cr
&\qquad\qquad\qquad
(1+\cos \chi)\,\,{\sin ^J}\chi\,\,
{C_{_{{\omega_1}-J-1}}^{^{J+1}}}(\cos \chi).\cr}
\eqno (42)
$$
\vskip 1pt
\centerline{
\epsfbox{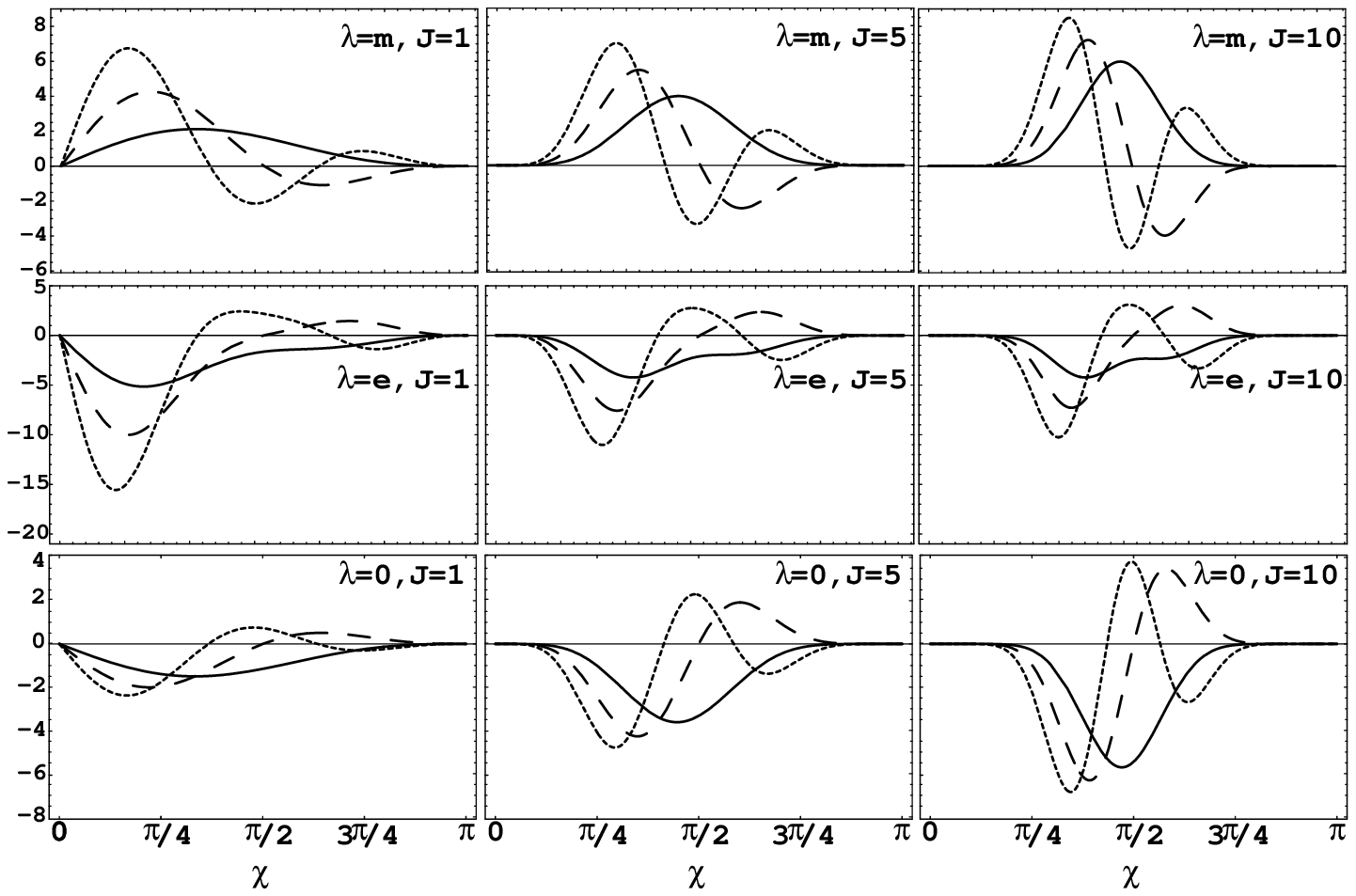}}
\vskip 30pt
\vbox{
\baselineskip=0.9\normalbaselineskip
\noindent
\figures {\caption Figure 3.}$\>\>$
From top to bottom, 
graphs of ${R_0^2}\,{{\cal I}_{_{\!J\!{\omega_1}}}^\lambda}(r)$ for
$\lambda=m,e,0$ and for different values of $J$ and $\omega_1$.
When $\lambda=m,0$, the solid line corresponds to 
${\omega_1}=J+1$, the dashed 
line correspond to ${\omega_1}=J+2$ and the dotted line
corresponds to ${\omega_1}=J+3$. For $\lambda=e$, the value of 
$\omega_1$ is equal to $J+2,\,J+3$ and $J+4$ for the solid line,
dashed line and the dotted line, respectively.
Note the different scales on the vertical axes.}
\vskip 20pt

Figure 3 shows the quantity 
${R_0^2}{{\cal I}_{_{\!J\!{\omega_1}}}^\lambda}(r)$
for different values of $J$ and $\omega_1$. 
It is important to
mention that because the index $({\omega_1}-J\pm 1)$ in
${C_{_{{\omega_1}-J\pm 1}}^{^{J+1}}}(\cos \chi)$ can only be
equal to zero or positive integers, the graphs of figure 3
contain degeneracies on $J$ and $\omega_1$. As shown in this
figure, for a given value of $R_0$, 
${{\cal I}_{_{\!J\!{\omega_1}}}^\lambda}(r)$ vanishes
at $\chi=0$ and $\pi$. This is quite expected since from
figure 2, the potential function of the Schr\"odinger-type
equation (35) approaches infinity at these two values of
$\chi$. This indicates that in a model closed FRW universe,
different modes of electromagnetic radiations, which 
correspond to different values of $\omega_1$, are
extended over the entire volume of the universe. 
Each of these modes is a member of a complete
set of eigenfunctions with amplitude that
increase by increasing the frequency.
An electromagnetic field is
a superposition of these eigenfunctions.
\vskip 10pt
{\centerline {\epsfbox{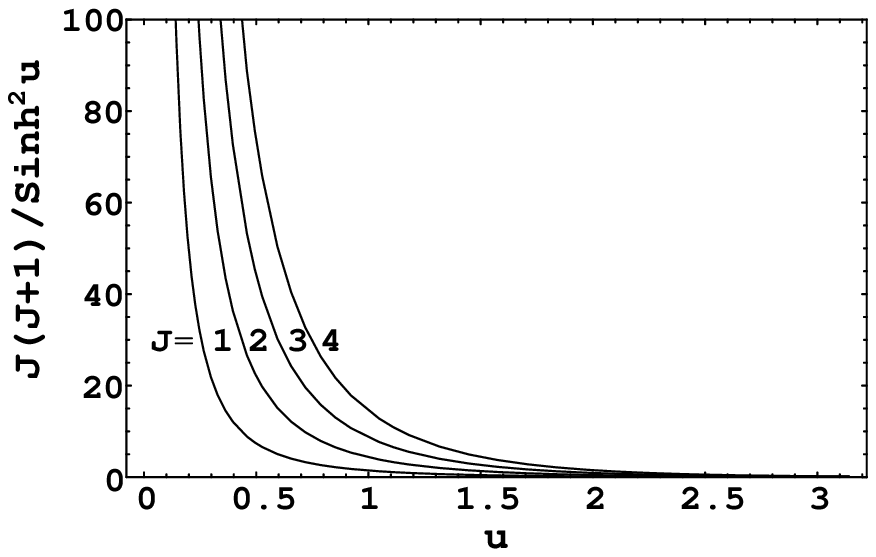}}}
\vbox{
\baselineskip=\normalbaselineskip
\vskip 60pt
\noindent
\figures{\caption Figure 4.}$\>\>$
Graph of $J(J+1)/\sinh^{2} u$ against $u$ for $J=1,2,3,4$.} 
\vskip 20pt
\noindent

{\timesboldeleven {3-c$\>$. Open Universe}}
\vskip 10pt

In an open FRW universe where $k=-1$, and
${f_k}(r)={f_{-1}}(r)=\qquad\qquad1/\{1-[r/(2{R_0})]^2\}$, 
equation  (22) can be written as 
\vskip 5pt
$$
{{d^2}\over {d{u^2}}}{\Phi_{_{\!J\!{\omega_{_{\!{-1}}}}}}^m}(u)\,+\,
\Biggl[{\omega_{_{\!{-1}}}^2}\,-\,
{{J(J+1)}\over{{\sinh^2}u}}\Biggr]\,
{\Phi_{_{\!J\!{\omega_{_{\!{-1}}}}}}^m}(u)\,=\,0\>.
\eqno  (43)
$$
\vskip 5pt
\noindent
In this equation, ${\omega_{_{\!{-1}}}}={R_0}{\sigma_{_{\!{-1}}}}$, 
$u=2{\tanh^{-1}}[r/(2{R_0})]$ and 
${\Phi_{_{\!J\!{\omega_{_{\!{-1}}}}}}^m}(u)=
{R_0}r\,{{\cal I}_{_{\!J\!{\sigma_{_{\!{-1}}}}}}^m}(r)$.
Equation (43) is a Schr\"odinger-type differential
equation with a {\it potential function} of
$J(J+1)/{\sinh^2}u$. Figure 4 shows this function for
different values of $J$.

With appropriate change of variables, it is possible to
show that equation (43) is, in fact, the differential
equation of a hypergeometric function. Let
\vskip 5pt
$$
{\zeta_{_{\!J\!{\omega_{_{\!{-1}}}}}}^m}(\xi)\,=\,
\big(i \csc\!{\rm h} u\big)^{i{\omega_{_{\!{-1}}}}}\,
{\Phi_{_{\!J\!{\omega_{_{\!{-1}}}}}}^m}(u),
\eqno (44)
$$
\vskip 5pt
\noindent
and
\vskip 5pt
$$
\xi \,=\,{1\over 2}\,(1-\coth u)\>.
\eqno (45)
$$
\vskip  5pt
\noindent
Substituting for
${\Phi_{_{\!J\!{\omega_{_{\!{-1}}}}}}^m}(u)$ and $u$ from equations
(44) and (45) in equation (43), one can show that this equation
can be written as
\vskip 8pt
\noindent
$$\eqalign{
\xi(1-\xi){{{d^2}
{\zeta_{_{\!J\!{\omega_{_{\!{-1}}}}}}^m}(\xi)}\over{d{\xi^2}}}
&+(1-i{\omega_{_{\!{-1}}}})(1-2\xi)
{{d{\zeta_{_{\!J\!{\omega_{_{\!{-1}}}}}}^m}(\xi)}\over {d\xi}}\cr
&+\Bigl[J(J+1)+i{\omega_{_{\!{-1}}}}(1-i{\omega_{_{\!{-1}}}})\Bigr]
{\zeta_{_{\!J\!{\omega_{_{\!{-1}}}}}}^m}(\xi)=0.\cr}
\eqno (46)
$$
\vskip 5pt
\noindent
Introducing
$
p\,=\,J+1-i{\omega_{_{\!{-1}}}},\,
q\,=\,-J-i{\omega_{_{\!{-1}}}}$, and
$s\,=1-i{\omega_{_{\!{-1}}}}$
equation (46) represents the differential equation of a 
hypergeometric function [22] with the general form of
\vskip 1pt
$$
\xi(1-\xi){{{d^2}
{\zeta_{_{\!J\!{\omega_{_{\!{-1}}}}}}^m}(\xi)}\over{d{\xi^2}}}
\,+\,\Bigl[s-(p+q+1)\,\xi\Bigr]\, 
{{d{\zeta_{_{\!J\!{\omega_{_{\!{-1}}}}}}^m}(\xi)}\over {d\xi}}\,
-\,p\,q\,{\zeta_{_{\!J\!{\omega_{_{\!{-1}}}}}}^m}(\xi)=0.
\eqno (47) 
$$
\vskip 5pt

Equation (47), as the general differential equation of
a hypergeometric function, has three regular singular
points at $\xi=0,\,1$ and $\infty$. In the neighborhood
of each of these singularities, equation (47) has two
independent solutions [22]. Since for all values of
$0<u<\infty$, from equation (45), $-\infty<\xi< 0$,
among all these solutions,
only the ones in the neighborhood of $\xi=0$ are valid 
solutions of equation (47).
Within this range, equation (47) has 
only one regular singular point at $\xi=0$, and its
general solution is given by [22]
$$
{\zeta_{_{\!J\!{\omega_{_{\!{-1}}}}}}^m}(\xi)=
{\cal B}{\cal F}(p,q,s,\xi)+{\cal D}{\xi^{1-s}}
{\cal F}(p-s+1,q-s+1,2-s,\xi).
\eqno (48)
$$
\vskip 1pt
\noindent
In this equation,
\vskip 1pt
$$
{\cal F}(p,q,s,\xi) = 
1+ {{pq} \over s}\,{\xi \over{1!}} +
{{pq}\over s}\,{{(p +1)(q +1)}\over {(s +1)}}\,
{{\xi^2}\over{2!}}+...\>,
\eqno  (49)
$$
\vskip 1pt
\noindent 
is the hypergeometric function, defined within its
circle of convergence $|\xi|<1$, and $\cal B$ and $\cal D$ are 
constant quantities. It is necessary to mention that
solution (48) converges conditionally
at $\xi=-1$, since ${\rm Rel}(s-p-q)=0$ [23].

To determine the values of the two constant quantities
$\cal B$ and $\cal D$, it is more convenient to study
the asymptotic character of 
${\Phi_{_{\!J\!{\omega_{_{\!{-1}}}}}}^m}(u)$ at
$u \to 0$ and $\infty$. Using the identity
\vskip 2pt
$$
{\cal F}(p,q,s,\xi)={(1-\xi)^{s-p-q}}{\cal F}(s-p,s-q,s,\xi),
\eqno(50)
$$ 
\vskip 2pt
\noindent
and considering that, as shown by equation (49), 
${\cal F}(p,q,s,\xi)$ is symmetric on $p$ and $q$,
equation (48) can be written as
\vskip 2pt
$$\eqalign{
{\zeta_{_{\!J\!{\omega_{_{\!{-1}}}}}}^m}(\xi)=
{\cal B}\,{(1-\xi)^{i{\omega_{_{\!{-1}}}}}}\,
&{\cal F}(-J,J+1,1-i{\omega_{_{\!{-1}}}},\xi)\cr
&+\,{\cal D}\,{\xi^{i{\omega_{_{\!{-1}}}}}}\,
{\cal F}(-J,J+1,1+i{\omega_{_{\!{-1}}}},\xi).\cr}
\eqno (51)
$$
\vskip 2pt
\noindent
From equations (44) and (45), one can show that
equation (51) simplifies to
$$\eqalign{
{\Phi_{_{\!J\!{\omega_{_{\!{-1}}}}}}^m}(u)=
{\cal B}\,\Bigl(-{{i\,{e^u}}\over 2}\Bigr)^{i{\omega_{_{\!{-1}}}}}\,
&{\cal F}(-J,J+1,1-i{\omega_{_{\!{-1}}}},\xi)\cr
&+\,{\cal D}\,\Bigl({{i\,{e^{-u}}}\over 2}\Bigr)^{i{\omega_{_{\!{-1}}}}}\,
{\cal F}(-J,J+1,1+i{\omega_{_{\!{-1}}}},\xi).\cr}
\eqno (52)
$$
\vskip 5pt
\noindent
As shown in figure 4, the potential function of the 
Schr\"odinger-type equation (43) approaches infinity
when $u \to 0$, and it tends to zero when $u \to \infty$. 
That implies, the solutions of equation (43) are finite
for all values of $u \in (0,\infty)$. 
The general solution should therefore be a superposition
of ingoing and outgoing radiation as in equation (52).
However, the physical interpretation of the FRW universes
based on the big bang model implies that all the radiation
must be outgoing. That means, when $u \to \infty$,  the
only possible solution for equation (43) is an out-going
wave [2,24]. This requires no in-coming radiation 
in equation (52). However, when $u \to \infty$, 
the quantity $\xi$ tends to zero, 
${\cal F}(p,q,s,\xi) \to 1$, and
the second term of equation (52)
represents a pure in-coming wave. The condition above
implies that, ${\cal D}=0$, and as a result,
the solution of equation (43) is written as
\vskip 2pt
$$
{\Phi_{_{\!J\!{\omega_{_{\!{-1}}}}}}^m}(u)=
{\cal B}\,\Bigl(-{{i\,{e^u}}\over 2}\Bigr)^{i{\omega_{_{\!{-1}}}}}\,
{\cal F}(-J,J+1,1-i{\omega_{_{\!{-1}}}},\xi).
\eqno (53)
$$
\vskip 5pt
\noindent
Setting ${\cal B}=1$, equation (53)
represents an out-going radiation with frequency 
${\omega_{_{\!{-1}}}}$. From the definition of 
${\Phi_{_{\!J\!{\omega_{_{\!{-1}}}}}}^m}(u)$, 
and from equations (20) and (21), the $r$-dependent
terms of electromagnetic fields are now given by
\vskip 10pt
$$\!\!\!\!\!\!\!\!\!\!\!\!\!\!\!\!\!\!\!\!\!\!
\!\!\!\!\!\!\!\!\!\!\!\!\!\!\!\!\!\!\!\!\!\!
\!\!\!\!\!\!\!\!\!\!\!\!\!\!\!\!\!\!\!\!\!\!
\!\!\!\!\!\!\!
{{\cal I}_{_{\!J\!{\omega_{_{\!{-1}}}}}}^m}(r)=
{1\over {r{R_0}}}\>{\cal X}(r)\,\,
{\cal F}\Bigl(-J,J+1,1-i{\omega_{_{\!{-1}}}},\xi (r)\Bigr),
\eqno (54)
$$
\vskip 5pt
$$\!\!
\eqalign{
&{{\cal I}_{_{\!J\!{\omega_{_{\!{-1}}}}}}^e}(r)=
-\,{i\over {r{R_0}}}\>{\cal X}(r)\,\,
\Biggl\{{\cal F}\Bigl(-J,J+1,1-i{\omega_{_{\!{-1}}}},\xi (r)\Bigr)\cr 
&\,\,+\,{{i\,{R_0^2}\,J(J+1)}\over
{2\,{r^2}\,{\omega_{_{\!{-1}}}}(1-i{\omega_{_{\!{-1}}}})}}\>
{\bigl[{f_{-1}}(r)\bigr]^{-2}}\>
{\cal F}\Bigl(-J+1,J+2,2-i{\omega_{_{\!{-1}}}},\xi (r)\Bigr) 
\Biggr\},\cr}
\eqno (55)
$$
\vskip 25pt
\noindent
and
$$\eqalign{
{{\cal I}_{_{\!J\!{\omega_{_{\!{-1}}}}}}^0}(r)=
-{1\over {{r^2}{\omega_{_{\!{-1}}}}}}
{\Bigl[J(J+1)\Bigr]^{1/2}}\,\,
&{\bigl[{f_{-1}}(r)\bigr]^{-1}}\cr
&{\cal X}(r)\,\,
{\cal F}\Bigl(-J,J+1,1-i{\omega_{_{\!{-1}}}},\xi (r)\Bigr).\cr} 
\eqno (56)
$$
\vskip 1pt
\centerline{
\epsfbox{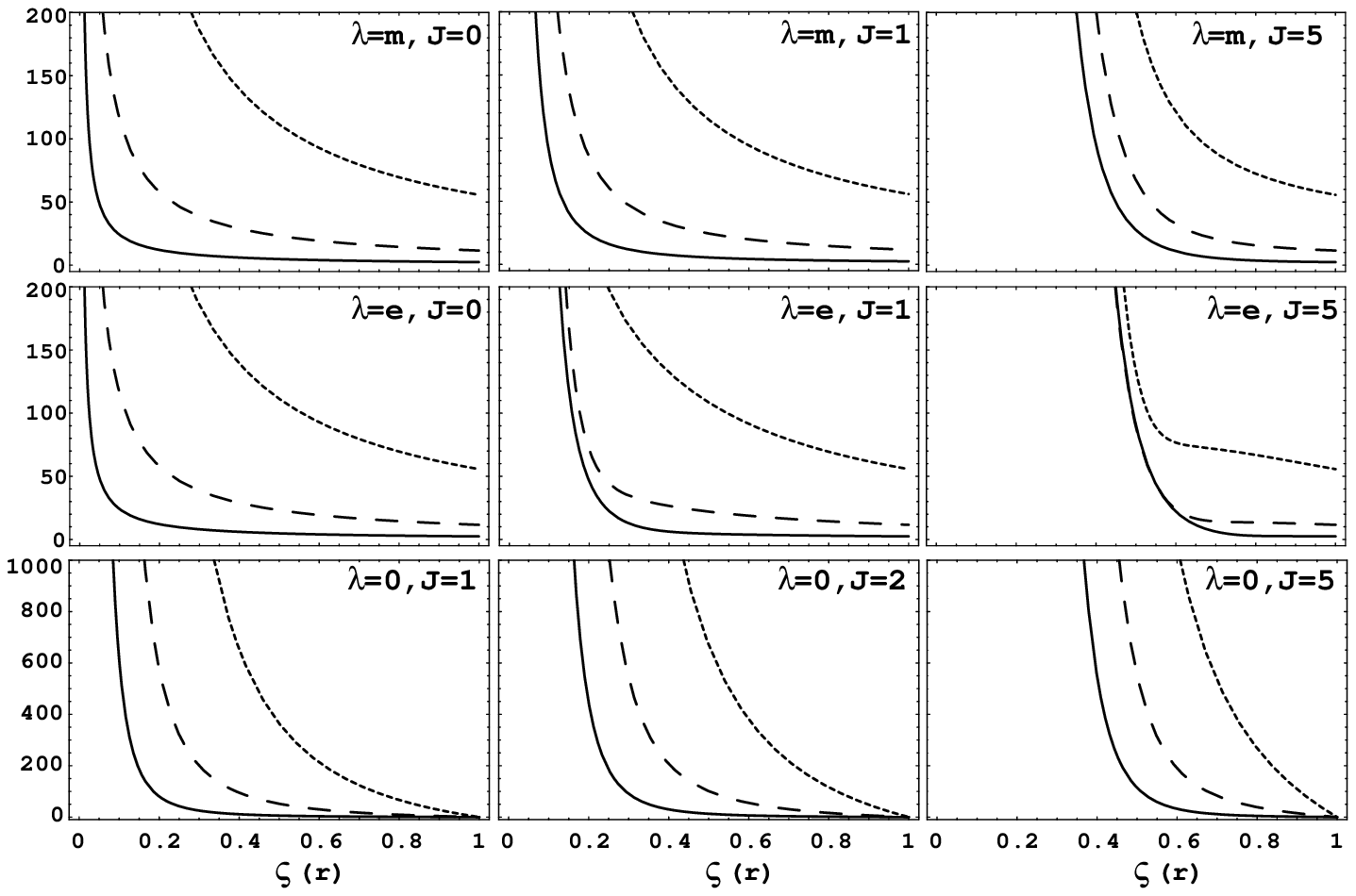}}
\vskip 30pt
\vbox{
\baselineskip=0.9\normalbaselineskip
\noindent
\figures {\caption Figure 5.}$\>\>$
From top to bottom, graphs of 
${R_0^2}|{{\cal I}_{_{\!J\!{\omega_{_{\!{-1}}}}}}^\lambda}(r)|$ for
$\lambda=m,e,0$, and for different values of $J$ and $\omega_{-1}$.
The solid line corresponds to ${\omega_{-1}}=1$, the dashed
line corresponds to ${\omega_{_{\!{-1}}}}=2$ and the dotted line 
corresponds to ${\omega_{_{\!{-1}}}}=3$. Note different scales on
vertical axes.}
\vskip 30pt
\noindent
In equations (54), (55), and (56), 
\vskip 2pt
$$
{\cal X}(r)={\Biggl[-{i\over 2}\,
{\biggl({{1+\varsigma (r)} \over {1-\varsigma (r)}}\biggr)}
\Biggr]^{i{\omega_{_{\!{-1}}}}}},
\eqno (57)
$$
\vskip 5pt
\noindent
$\varsigma(r)=({r/{2{R_0}}})$,
and the derivatives of the hypergeometric function
${\cal F}(p,q,s,\xi(r))$ have been replaced by
\vskip 1pt
$$
{d\over {dr}}\,{\cal F}{\Bigl(}p\,,\,q\,,\,s\,,\,\xi (r){\Bigr)}\,=\,
{{pq}\over s}\,{\cal F}\Bigl(p+1\,,\,q+1\,,\,s+1\,,\,\xi(r)\Bigr)
\biggl[{{d\xi(r)}\over {dr}}\biggr].
\eqno (58)
$$
\noindent
Figure 5 shows the product of the absolute values of 
complex functions
${{\cal I}_{_{\!J\!{\omega_{_{\!{-1}}}}}}^\lambda}(r)$ 
and $R_0^2$ in terms of $\varsigma$, and
for different values of $J$ and ${\omega_{_{\!{-1}}}}$.
It is necessary to mention that in an open FRW universe,
${\cal R}\in (0,\infty)$, which from equation (13) 
\vskip 1pt
\centerline{
\epsfbox{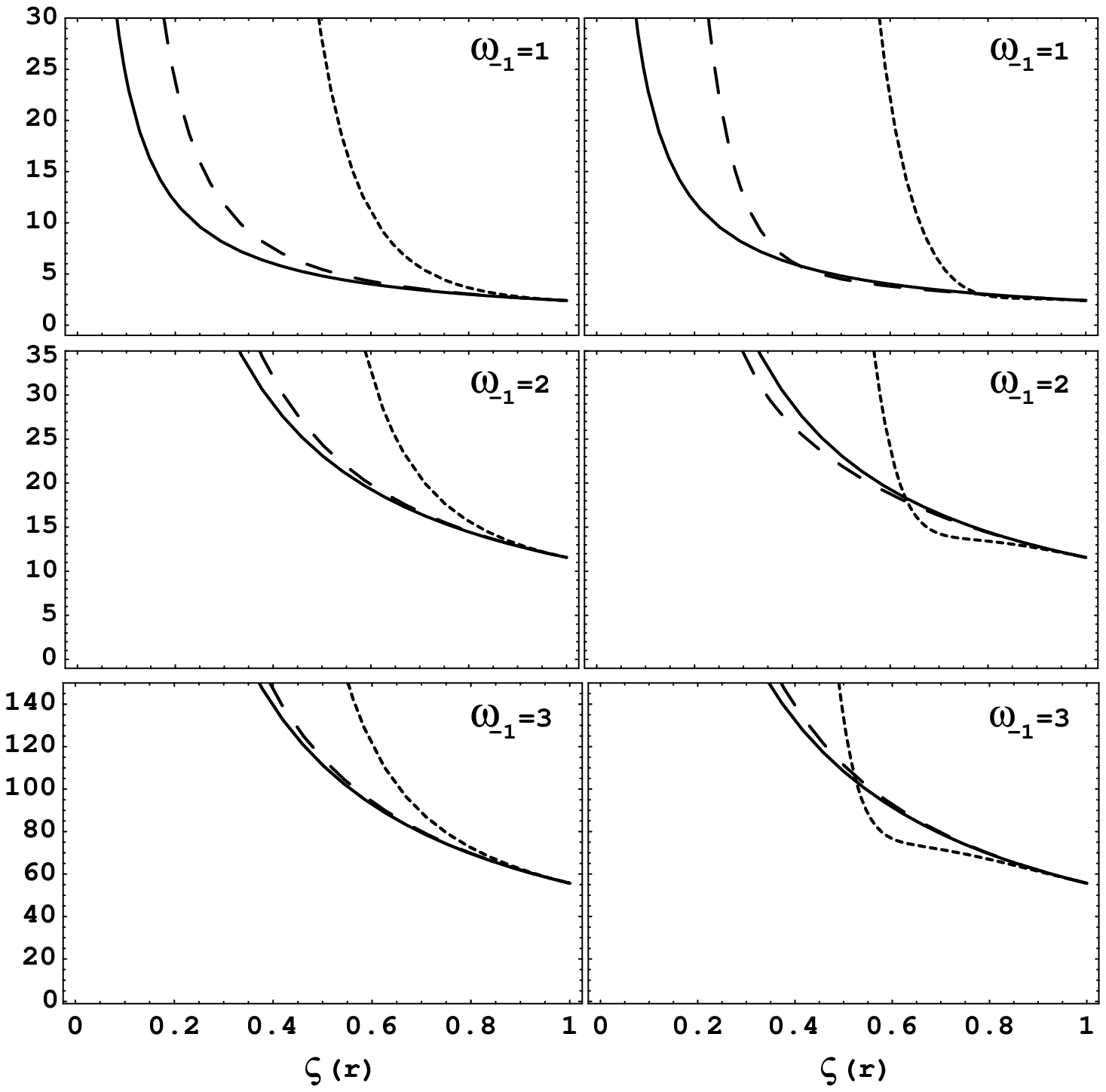}}
\vskip 2.2in
\vbox{
\baselineskip=0.9\normalbaselineskip
\noindent
\figures {\caption Figure 6.}$\>\>$
Graphs of 
${R_0^2}|{{\cal I}_{_{\!J\!{\omega_{_{\!{-1}}}}}}^\lambda}(r)|$ for
$\lambda=m$ (left column), $\lambda=e$ (right column), 
and for different values of $J$ and $\omega_{-1}$.
The solid line corresponds to $J=0$, the dashed
line corresponds to $J=1$ and the dotted line 
corresponds to $J=5$. As mentioned in the text, the quantities 
${R_0^2}|{{\cal I}_{_{\!J\!{\omega_{_{\!{-1}}}}}}^m}(2{R_0})|$
and ${R_0^2}|{{\cal I}_{_{\!J\!{\omega_{_{\!{-1}}}}}}^e}(2{R_0})|$
are independent of the value of $J$ and are equal to
$0.5\,{\rm exp}({\omega_{_{\!{-1}}}}\pi/2)$. For
${\omega_{_{\!{-1}}}}=1$ this value is $\sim 2.4$ (top graphs),
for ${\omega_{_{\!{-1}}}}=2$, it is $\sim 11.57$ (middle graphs),
and when ${\omega_{_{\!{-1}}}}=3$, it attains the value of
$\sim 55.66$ (bottom graphs). Note different scales on vertical axes.}
\vskip 50pt
\noindent
implies
$0<r< 2{R_0}$. As shown here, for a given value of the radius
of universe, ${{\cal I}_{_{\!J\!{\omega_{_{\!{-1}}}}}}^\lambda}(r)$ 
approaches zero for $\lambda=0$, and a non-zero
constant value for $\lambda=e,m$.
Using identities
${X^Y}={\rm exp}(Y{\rm Ln}X)$, and 
$\pm i = {\rm exp}[\pm i\,({\pi/2}+2\pi l)]$, 
and choosing the principal branch on the complex plane,
one can show that this constant value is equal to
$0.5\,{\rm exp}({\omega_{_{\!{-1}}}}\pi/2)$ 
as demonstrated in figure 6. 
\vskip 20pt
\noindent
{\timesboldeleven {4. $\>$ Asymptotic Character of
${\vec E} (\vec r,\eta)$ and ${\vec H} (\vec r,\eta)$}}
\vskip 10pt
To study the asymptotic behavior of electromagnetic 
fields, one has to evaluate the magnitude of $\vec E$ 
and $\vec H$ for large values of $\cal R$. To do so,
imagine a set of fundamental observers in FRW universes,
at rest in space with constant spatial coordinates
${x_1},\,{x_2}$ and $x_3$ as given
by equation (11) . At these coordinates, 
the measured components of the Faraday tensor are given by
$$
{F_{\!(\rho)\!(\nu)}}\,=\,{F_{\alpha\beta}}\,
{{\Lambda^\alpha}_{\!(\rho)}}{{\Lambda^\beta}_{\!(\nu)}}\>,
\eqno  (59)
$$
\noindent
where
$$\!\!\!\!\!\!\!\!\!\!\!\!\!\!\!\!\!\!
{{\Lambda^\alpha}_{(0)}}\,=\,
\Bigl({{\cal C}^{-1}}(\eta)\,,\,0\,,\,0\,,0\Bigr),
\eqno  (60)
$$
$$
{{\Lambda^\alpha}_{(1)}}\,=\,
\Bigl(0\,,{{\cal C}^{-1}}(\eta){f^{-1}_k}(r)\,,\,0\,,\,0\Bigr),
\eqno  (61)
$$
$$
{{\Lambda^\alpha}_{(2)}}\,=\,
\Bigl(0\,,0\,,{{\cal C}^{-1}}(\eta){f^{-1}_k}(r)\,,\,0\Bigr),
\eqno  (62)
$$
$$
{{\Lambda^\alpha}_{(3)}}\,=\,
\Bigl(0\,,0\,,0\,,{{\cal C}^{-1}}(\eta){f^{-1}_k}(r)\Bigr),
\eqno  (63)
$$
\noindent
are the orthonormal tetrad frames associated with these observers
calculated from equation
$$
{g^{\alpha\beta}}\,=\,{\eta_{_M}^{(\rho)\!(\nu)}}\,
{{\Lambda^\alpha}_{\!(\rho)}}{{\Lambda^\beta}_{\!(\nu)}}\>.
\eqno (64)
$$
\vskip 1pt
\noindent
In equation (64), 
$\eta_{_M}^{(\rho)(\nu)}$ is the metric of the Minkowski spacetime
and $g^{\alpha\beta}$ represents the line element of FRW universes
as given by equation (11). 
Using equations (3) and (4) and also equations (7) to (9), the
components of Faraday tensor ${F}_{\!(\rho)\!(\nu)}$ can be written as
$$
{F_{\!(\rho)\!(\nu)}}\,=\,{{\cal C}^{-2}}(\eta)\,
{\bigl[{f_k}(r)\bigr]^{-1}}\,
\pmatrix{ 0&    {E_1}&   {E_2}&    {E_3}\cr
         -{E_1}&        0&     {H_3}&   -{H_2}\cr
         -{E_2}&     -{H_3}&     0&      {H_1}\cr
         -{E_3}&      {H_2}&  -{H_1}&     0  \cr}\>.
\eqno (65)
$$
\vskip 2in

In equation (65), $({E_1},{E_2},{E_3})$ and $({H_1},{H_2},{H_3})$
represent the components of the electric and magnetic fields 
${\vec E} ({\vec r},\eta)$ and ${\vec H} ({\vec r},\eta)$, as given
by equations (23) and (24), along $x_1$- $x_2$- and $x_3$-axis.
To study the asymptotic behavior of $\vec E$ and $\vec H$, one
has to study the properties of these components for large values
of $\cal R$. Such a study is meaningful only
within the context of a flat and an open FRW universe.
In a closed FRW universe, as shown in section 3-b, at any
given epoch, the radial parts of the solution of equations 
(15) and (16) extend within the entire volume of the universe.
This characteristic of 
${{\cal I}_{_{\!J\!{\omega_{_{\!{1}}}}}}^\lambda}(r)$ 
makes the study of the asymptotic behavior 
of electromagnetic fields in a closed FRW universe implausible. 

From equations (18) and (19), and for given values of
$J, M, {\sigma_k}$ and $\lambda$, the $r$-dependence of 
electromagnetic fields
is given by ${\cal I}_{\!J\!{\sigma_k}}^\lambda (r)$. 
As a result, the measured Faraday tensor
$F_{\!(\rho)\!(\nu)}$ will have an $r$-dependence of the form
${\cal I}_{\!J\!{\sigma_k}}^\lambda (r)/{f_k}(r)$. 
In a flat FRW universe, ${\cal R}={r/{R_0}}\,,{f_0}(r)=1$,
and from equations (31), (33) and (34), for a constant
value of $\sigma_0$, the functions
${\cal I}_{\!J\!{\sigma_k}}^\lambda (r)$ are proportional
to spherical Bessel functions ${j_{_J}}(r)$. For large values
of $\cal R$, the asymptotic values of these functions are
given by [22,25]
\vskip 1pt
$$
{j_{_J}}(r')\sim {1\over r'}\sin\Bigl(r'-{{J\pi}\over 2}\Bigr).
\eqno (66)
$$
\vskip 1pt
\noindent
Equation (66) indicates that, as expected, in a flat FRW universe, 
electromagnetic fields approach zero at large distances
with amplitudes decreasing as ${\cal R}^{-1}$. A result that
is also seen from figure 1.
In an open FRW, as indicated by equations (54), (55) and (56),
${\cal I}_{\!J\!{\omega_{_{\!{-1}}}}}^\lambda (r)$ 
approach constant values as $r\to 2{R_0}$.
Therefore from equation (65), the measured fields approach
zero since ${[{f_{-1}}(r)]^{-1}}=\{1-[r/(2{R_0})]^2\}\to 0$. It follows
that, similar to a flat universe, in an open FRW universe also, 
measured fields approach zero as
${\cal R}^{-1}$ as ${\cal R}\to \infty$.
\vskip 20pt
{\centerline{\timesboldeleven {5. SUMMARY}}}
\vskip  5pt

The results of a study of the
asymptotic properties of electromagnetic waves in a FRW 
universe were presented. Electromagnetic fields were 
considered as small perturbations on the background curvature 
of FRW spacetime, and the metric of the universe was written
in form of a line element that, at a conformal time
corresponding to a given epoch, is electromagnetically equivalent
to FRW metric. The solutions of Maxwell's equations were 
obtained for all three cases of flat, 
closed and open FRW universes.

The isotropy and homogeneity of FRW metric allows
writing the solutions of the electromagnetic field equations 
in terms of vector spherical harmonics. 
Expansions to generalized spherical harmonics have
also been presented by Mankin et al [13-15,26], and
by Laas et al [27] to construct
exact and approximate solutions to electromagnetic
wave equations in a curved spacetime 
using higher-order Green's function method.

It was shown that using appropriate transformations, 
the equations governing the $r$-dependence of
the electromagnetic fields for the type-$m$ vector spherical harmonics
could be written in form of a one-dimensional 
Schr\"odinger-type equation
whose eigenvalues represent the frequencies of 
different modes of electromagnetic radiations. 
In a closed FRW universe, the solutions of the
above-mentioned equation resemble
standing waves, extended over the entire volume of the universe,
implying that the study of their asymptotic character would be
implausible.

The asymptotic values of electromagnetic fields
in a flat and an open FRW universe were obtained
by calculating these fields in the local frame of an observer
at large distances. Analysis of the results indicated that,
as expected, these fields tend to zero as ${\cal R}^{-1}$ 
at large distances.

An application of the results of the analysis presented here would be
in the study of the tails of
electromagnetic waves in FRW universes.  As shown by Noonan [5,6,7],
the vector potential of an electromagnetic wave has a non-zero tail when
propagating in a curved spacetime. However, in conformally flat
universes such as FRW, because of the conformal invariance of
Maxwell's equations, the components of Faraday's tensor have no
tails [5,6]. Such a conclusion can trivially be made in a 
flat universe where the background spatial curvature is non-existent,
and can be attributed to the fact that in a (3+1) flat universe,
the four dimensional Green's function of the spacetime
has a delta-function character, and as a result,
only motions along the light cone are supported.
For a closed FRW universe, as mentioned in section 3-b,
because the solutions of its corresponding 
Schr\"odinger-type differential equation resemble standing waves,
the concept of tails is not applicable. 
However, such studies are quite plausible in an open FRW universe.
For a specific electromagnetic wave, in this case, 
the tail-free nature of the components of Faraday tensor
can be examined by expanding these components
in terms of
${\vec E}_{\!\!J\!M\!{\sigma_k}}^\lambda (\vec r\,)$ and
${\vec H}_{\!\!J\!M\!{\sigma_k}}^\lambda (\vec r\,)$ as
given by equations (18) and (19),  and calculating their
asymptotic values for large values of $\cal R$. 
For given values of $J\,,M$ and $\sigma_0$, the 
$r$-dependence of these fields are given by
${{\cal I}_{\!J\!{\omega_{_{\!{-1}}}}}^\lambda}\!(r)/{f_{-1}}(r)$.
As shown in section 4, for large values of $\cal R$,
these quantities approach zero as ${\cal R}^{-1}$, indicating the
tail-free nature of electromagnetic waves.
\vskip  20pt
\noindent
{\centerline {\timesboldeleven {ACKNOWLEDGEMENT}}}
\vskip 5pt
I am indebted to Bahram Mashhoon for critically reading 
the original manuscript and also for his valuable 
suggestions and comments.

\vskip 50pt
\noindent
{\timesboldeleven {REFERENCES}}
\vskip 20pt
\noindent
\hskip 5pt
[1] DeWitt, B. S., and Brehme, R. W. (1965).
{\it Ann. Phys. (NY)} {\bf 9}, 220.
\vskip 2pt
\noindent
\hskip 5pt
[2] Faraoni, V., and Sonego, S. (1992). {\it Phys.Let. A} 
{\bf 170}, 413.
\vskip 2pt
\noindent
\hskip 5pt 
[3] Sonego, S., and Faraoni, V. (1992). {\it J.Math.Phys.} 
{\bf 33}, 625.
\vskip 2pt
\noindent 
\hskip 5pt 
[4] Scialom, D., and Philippe, J. (1995). {\it Phys.Rev.D}
{\bf 51}, 5698.
\vskip 2pt
\noindent
\hskip 5pt
[5] Noonan, W. T. (1989). {\it Astrophys. J.} {\bf 341}, 786.
\vskip 2pt
\noindent
\hskip 5pt 
[6] Noonan, W. T. (1989). {\it Astrophys. J.} {\bf 343}, 849.
\vskip 2pt
\noindent 
\hskip 5pt 
[7] Noonan, W. T. (1995). {\it Class. Quantum Gravit.}
{\bf 12}, 1087.
\vskip 2pt
\noindent
\hskip 5pt 
[8] Schr\"odinger, E. (1939). {\it Physica} VI, {\bf 9}, 899.
\vskip 2pt
\noindent
\hskip 5pt 
[9] Schr\"odinger, E. (1940). {\it Proc.R.I.A.} {\bf IV}, 23.
\vskip 2pt
\noindent
[10] Infield, L., and Schild, A. E. (1945). {\it Phys.Rev.} 
{\bf 68}, 250.
\vskip 2pt
\noindent
[11] Infield, L., and Schild, A. E. (1946).
{\it Phys.Rev} {\bf 70}, 410.
\vskip 2pt
\noindent
[12] Deng, Y., and Mannheim, P.D. (1988). 
{\it Gen. Relative Gravit.} {\bf 20}, 969.
\vskip 2pt
\noindent
[13] Mankin, R., Laas, T., and Tammelo. R. (2001).
{\it Phys.Rev.D}, {\bf 63}, 063003.
\vskip 2pt
\noindent
[14] Mankin, R., Tammelo. R., and  Laas, T. (1999).
{\it Class. Quantum Grav.},
\vskip 2pt
\noindent
\hskip 18pt
{\bf 16}, 1215.
\vskip 2pt
\noindent
[15] Mankin, R., Tammelo. R., and Laas, T. (1999).
{\it Class. Quantum Grav.},
\vskip 2pt
\noindent
\hskip 18pt
{\bf 16}, 2525.
\vskip 2pt
\noindent
[16] Hadamard, J. (1923). {\it Lectures on Cauchy's Problem}
(Yale Univ.
\vskip 2pt
\noindent
\hskip 18pt
Press, New Haven, CT).
\vskip 2pt
\noindent
[17] Skrotskii, G. V. (1957). {\it Sov.Phys.-Dokl.} {\bf 2}, 226.
\vskip 2pt
\noindent
[18] Plebanski, J. (1960). {\it Phys.Rev.} {\bf 118}, 1396.
\vskip 2pt
\noindent
[19] Mashhoon, B. (1973). {\it Phys.Rev.D} {\bf 8}, 4297.
\vskip 2pt
\noindent
[20] Newton, R. G. (1966). {\it Scattering Theory of Waves and Particles}
\vskip 2pt
\noindent
\hskip 18pt
(McGraw-Hill, New York).
\vskip 2pt
\noindent
[21] Davydov, A. S. (1967). {\it Quantum Mechanics}
(NEO Press, Ann Arbor, 
\vskip 2pt
\noindent
\hskip 18pt
Michigan). 
\vskip 2pt
\noindent
[22] Magnus, W., Oberhettinger, F., and Soni, R. P. (1966).
{\it Formulas and 
\vskip 2pt
\noindent
\hskip 18pt
Theorems for the Special Functions of Mathematical Physics}
\vskip 2pt
(Springer, New York). 
\vskip 2pt
\noindent
[23] Abramowitz, M., and Stegun, I. A. (1964).
{\it Handbook of Mathematical
\vskip 2pt
Functions} (National Bureau of Standards, Washington, DC).
\vskip 2pt
\noindent
[24] Kundt W., and Newman, E. T. (1968). {\it J. Math. Phys.} 
{\bf 9}, 2193.
\vskip 2pt
\noindent
[25] Arfken, G. A. (1985). {\it Mathematical Methods for
Physicists} 
\vskip 2pt
(Academic Press, New York).
\vskip 2pt
\noindent
[26] Mankin, R., Tammelo. R. and  Laas, T. (1999).
{\it Gen. Relativ. Gravit.},
\vskip 2pt
\noindent
\hskip 18pt
{\bf 31}, 537.
\vskip 2pt
\noindent
[27] Laas, T., Mankin, R., and Tammelo. R. (1998).
{\it Class. Quantum Grav.},
\vskip 2pt
\noindent
\hskip 18pt
{\bf 15}, 1595.

\bye